\documentclass[sigconf, nonacm]{acmart}
\AtBeginDocument{%
  }

\setcopyright{acmlicensed}
\copyrightyear{2018}
\acmYear{2018}
\acmDOI{XXXXXXX.XXXXXXX}
\acmConference[Conference acronym 'XX]{Make sure to enter the correct
  conference title from your rights confirmation email}{June 03--05,
  2018}{Woodstock, NY}
\acmISBN{978-1-4503-XXXX-X/2018/06}

\begin{document}

\title{PagedMoE: Cross-Precision Expert Caching for Token-Level Dynamic Precision in SSD-Resident MoE Inference}

\author{Yutong Lin}
\orcid{0009-0009-4064-8032}
\affiliation{%
  \institution{Tsinghua University}
  \city{Beijing}
  \country{China}
}
\email{lin-yt21@mails.tsinghua.edu.cn}

\author{Yingdi Shan}
\correspondingauthor
\affiliation{%
  \institution{Tsinghua University}
  \city{Beijing}
  \country{China}
}
\email{shanyd@tsinghua.edu.cn}

\author{Shaoyuan Chen}
\affiliation{%
  \institution{Tsinghua University}
  \city{Beijing}
  \country{China}
}
\email{chensy20@mails.tsinghua.edu.cn}

\author{Yuanyong Chen}
\affiliation{%
  \institution{Tsinghua University}
  \city{Beijing}
  \country{China}
}
\email{yy-chen24@mails.tsinghua.edu.cn}

\author{Yongwei Wu}
\affiliation{%
  \institution{Tsinghua University}
  \city{Beijing}
  \country{China}
}
\email{wuyw@tsinghua.edu.cn}

\author{Mingxing Zhang}
\correspondingauthor
\affiliation{%
  \institution{Tsinghua University}
  \city{Beijing}
  \country{China}
}
\email{zhang\_mingxing@mail.tsinghua.edu.cn}

\renewcommand{\shortauthors}{Lin et al.}

\begin{abstract}
Large Mixture-of-Experts (MoE) models are promising for local inference because sparse routing allows the routed-expert pool to reside on SSD while DRAM caches recently used experts. However, under practical DRAM cache capacities, substantial SSD I/O remains on the critical path despite I/O--computation overlap. Token-level dynamic expert precision offers a natural way to reduce this data demand by allocating bits according to expert importance and quantization sensitivity. Nevertheless, treating each bit-width of the same expert as an independent object prevents cross-precision cache reuse, causing overlapping data to be fetched repeatedly.

The key insight is that precision should instead denote the refinement depth of one shared expert state. Based on this insight, we present PagedMoE, which stores each expert once using a prefix-compatible tree codebook and bitplane layout. Lower-precision representations are valid prefixes of higher-precision ones, allowing cached data to be reused or extended by fetching only missing bitplanes. PagedMoE selects the required prefix length before I/O using current routing weights and offline expert-specific error profiles, and directly executes cached 1--8-bit prefixes without materializing full-precision experts.

At an average routed-expert precision of 2.5 bits, PagedMoE achieves quality close to GPTQ-INT4 on Qwen3.5-397B-A17B and to MXFP4 on most evaluated GLM5.2 tasks. With a 24~GiB expert cache, it reaches 8.89~tokens/s on Qwen3.5, \(1.33\times\) the throughput of the same dynamic requests using independent precision objects and \(1.56\times\) that of uniform 4-bit execution.
\end{abstract}




\setcopyright{none}

\settopmatter{
  printacmref=false,
  printccs=false,
  printfolios=true
}


\maketitle

\section{Introduction}

Large Mixture-of-Experts (MoE) models combine a large parameter
capacity with sparse per-token computation. A routed MoE layer may
contain hundreds of experts, while each token activates only a small
subset of them. This sparsity makes batch-size-one local inference
possible even when the complete model cannot remain in GPU memory or
host DRAM. Dense operators and attention can execute on the GPU,
selected routed experts can execute on the CPU, and an SSD can provide
capacity for the complete expert pool. Host DRAM then acts as a cache
for recently used expert weights~\cite{ktransformers,fiddler,moeinfinity}.

In this architecture, however, sparse computation does not imply
sparse data movement. Successive tokens select different experts, so
the complete expert pool must remain accessible. Temporal locality
makes expert caching effective, but practical cache capacities do not
eliminate SSD access. On Qwen3.5-397B-A17B, our optimized fixed-4-bit
runtime serves 57.08\% of requested expert bytes from a 24~GiB DRAM
cache, yet still reads 1.53~GiB from SSD per generated token. Even
after overlapping asynchronous I/O with expert computation,
87.16~ms/token of SSD-related stall remains exposed. This stall
accounts for 49.28\% of decode latency and limits throughput to
5.65~tokens/s. Thus, after caching and I/O--computation overlap have
been optimized, the amount of expert data requested by each token
remains a first-order bottleneck.

Reducing expert precision lowers both the bytes transferred on a cache
miss and the DRAM footprint of a resident expert. Uniform quantization,
however, assigns the same precision to every selected expert for every token.
This ignores two forms of variation for MoE layers: routing weights vary across
tokens, and experts obtain different quality benefits from additional
precision. These variations create an opportunity to allocate a fixed bit budget nonuniformly across the experts selected for each token.

The central systems problem is that dynamic precision is not
automatically cacheable. A straightforward implementation stores each
supported precision as an independent expert object, giving different
bit-widths of the same expert different cache identities. A cached
2-bit object then cannot contribute to a later 4-bit request, even on
consecutive tokens. Dynamic precision can therefore reduce
per-invocation bytes while destroying reuse across invocations. At a
24~GiB cache capacity, fixed 4-bit requests achieve a 57.08\%
exact-object hit rate on our trace, whereas dynamic requests with the
same average 4-bit budget achieve only 35.15\% when each precision is
treated as an independent object. The problem is therefore not
dynamic precision itself, but representing precision as a set of
mutually disjoint states.

This loss of reuse is not inherent to dynamic precision. All supported
bit widths approximate the same underlying weights, yet conventional
formats encode those approximations independently. The key observation
is that the precision levels can instead be constructed as a nested
sequence: moving from $b$ to $b+1$ bits adds refinement information to
a coarser reconstruction. Exposing this nesting to the storage
hierarchy allows a resident low-precision state to become the reusable
prefix of a later higher-precision request, while a resident
high-precision state can serve a lower-precision request directly.

This observation suggests a design principle: \emph{Precision should identify the refinement depth of one expert state, not the version of an expert object}. Guided by this principle, we
design PagedMoE. Each expert is stored once as an ordered sequence of
refinement units, and its precision is determined by the
length of the prefix used for execution. If a cached expert currently
has an $x$-bit prefix and a later request requires $b>x$ bits, the
runtime loads only the missing suffix. If $b\leq x$, the resident
state directly serves the request. All precision levels therefore
share one SSD-resident representation, while precision changes reuse
DRAM residency and preserve the work already performed to bring an
expert into memory.

Prior work has established that a single checkpoint can expose
multiple bit widths~\cite{park2024anyprecision,nair2025matquant} and
that MoE systems can vary expert precision according to runtime
importance~\cite{hobbit,slicemoe,dymoe,d2moe}. 
PagedMoE builds on these directions but addresses a distinct
storage-hierarchy requirement: expert states at fine-grained precision levels must remain
incrementally reusable from SSD through the DRAM cache, and directly consumable by execution, without
materializing a dequantized expert copy.
Its contribution is not the existence of multiple precisions, but an
end-to-end design in which a precision change either extends one
cached state or uses a shorter prefix of that state, rather than
selecting a different expert object.

Making this abstraction practical requires three properties across the
execution path. First, every prefix must reconstruct a valid expert,
and each additional refinement must be independently addressable on
SSD and in DRAM. Second, the target bit width must be chosen after routing
but before expert loading, using only information available before the
requested representations are fetched; otherwise, the system cannot
know which refinements to load. Third, the selected prefix must be
executed directly from its compact cached form. Reconstructing a complete BF16 expert or staging a separate compute-oriented representation would introduce additional memory traffic,
negating much of the data-movement benefit of compact storage and
caching.

PagedMoE realizes these properties through a co-designed
representation, allocator, and runtime. A tree-structured codebook
makes every prefix of a quantization index a valid reconstruction, and
a bitplane layout stores each additional index bit as an independently
loadable refinement unit. A pre-I/O allocator combines current routing
weights with offline expert-specific output-error profiles to assign a
target bit width to each selected expert under an average requested-bit
budget. The DRAM cache maintains expert prefixes, loads only missing
bitplanes, and evicts only suffixes. Finally, an AVX-512 CPU kernel
reconstructs values from the requested bitplanes on the fly and
immediately consumes them in GEMV, without materializing a
full-precision expert.

Our evaluation separates the benefit of requesting fewer bits from
that of cross-precision reuse. On Qwen3.5-397B-A17B, with a 24~GiB
expert cache, uniform 4-bit execution reaches 5.71~tokens/s. Dynamic
Avg2.5 without cross-precision reuse reaches 6.67~tokens/s, showing the
benefit of requesting fewer bits alone. With the same expert requests and bit-width assignments, PagedMoE
represents each expert's precision levels as prefixes of one shared
progressive state and reaches 8.89~tokens/s, \(1.33\times\) that of
the independent configuration and \(1.56\times\) that of uniform
4-bit execution. GLM5.2 shows the same trend. At an average
routed-expert precision of 2.5 bits, PagedMoE remains close to
GPTQ-INT4 on the evaluated Qwen3.5 tasks and to MXFP4 on most evaluated
GLM5.2 tasks, although the quality cost is model- and task-dependent.
In a complete-runtime comparison on Qwen3.5 under total-memory
accounting, PagedMoE achieves \(3.58\times\) the throughput of
\texttt{ik\_llama.cpp} while using less host memory.

This paper makes the following contributions:

\begin{itemize}

    \item \textbf{A cache-centric formulation of dynamic expert
    precision.}
    We show that reducing requested precision is insufficient when
    different bit widths are represented as independent objects:
    precision-specific objects can erase much of the locality available in
    the expert access stream. We formulate precision instead as the
    refinement depth of one shared, monotonically extendable expert
    state.

    \item \textbf{An end-to-end representation and runtime for
    cross-precision reuse.}
    PagedMoE combines prefix-compatible tree-codebook quantization,
    independently addressable bitplanes, a prefix-preserving DRAM
    cache, incremental asynchronous SSD loading, and direct CPU
    execution. The same compact state is preserved across storage,
    caching, refinement, and computation.

    \item \textbf{Pre-I/O allocation and an evaluation that separates
    its system effects.}
    PagedMoE assigns per-token, per-expert bit-widths from
    routing weights and offline error profiles. Experiments on two
    large MoE models quantify model quality, the benefit of requesting
    fewer bits, the additional benefit of cross-precision reuse, and
    the importance of preserving compactness through execution.

\end{itemize}

\section{Background and Motivation}

\subsection{SSD-Resident Local MoE Inference}
\label{sec:ssd-resident-moe}

A routed Mixture-of-Experts (MoE) layer contains a pool of experts but
evaluates only a small subset for each token. Given an input
representation \(x\), the router selects \(K\) experts and assigns them
normalized routing weights \(g_1,\ldots,g_K\). The routed output is

\begin{equation}
    y = \sum_{i=1}^{K} g_i E_{e_i}(x),
    \label{eq:routed-output}
\end{equation}

where \(e_i\) is the \(i\)-th expert selected for the current token
and \(E_{e_i}\) is its feed-forward function. Expert identity is
layer-specific.

Sparse routing substantially reduces per-token computation but not the
capacity required for the routed-expert pool.
Qwen3.5-397B-A17B, for example, activates only 10 of 512 experts per
MoE layer, totaling 7.55 billion parameters per token,
compared with 386.55 billion parameters in its complete routed-expert
pool~\cite{qwen35_397b}. Because subsequent tokens may select any expert in each layer,
the complete pool must remain accessible.

We consider batch-size-one local decoding on a heterogeneous machine.
Dense layers, attention, routing, and other always-activated operators
reside and execute on the GPU. The complete routed-expert pool resides
on SSD, selected experts execute on the CPU, and host DRAM caches expert
data between SSD storage and CPU execution. This organization makes
models larger than GPU and host memory executable on a local PC.

Decode throughput then depends on the number of expert bytes requested,
the fraction already resident in DRAM, the SSD latency exposed after
I/O--computation overlap, and the DRAM traffic required to execute
resident experts. Caching can reduce cache misses, and asynchronous
execution can hide part of the remaining I/O latency, but neither
reduces the payload of a non-resident expert request.

\subsection{Fixed-Precision Caching Leaves SSD I/O on the Critical Path}
\label{sec:practical-cache-operating-point}

Routed-expert accesses exhibit temporal locality during batch-size-one
decoding. A DRAM cache can therefore satisfy many expert requests
without accessing SSD. 
The key question is whether practical cache capacities capture enough
reuse to remove SSD I/O from the decoding critical path. To answer this
question, we implement an optimized fixed-4-bit configuration, denoted
\emph{Fixed4}. It caches each complete 4-bit expert representation as
one object, uses a fused low-bit CPU kernel, and overlaps asynchronous
SSD reads with cache-ready expert computation.

\begin{figure}[t]
\centering
\includegraphics[width=0.9\linewidth]
{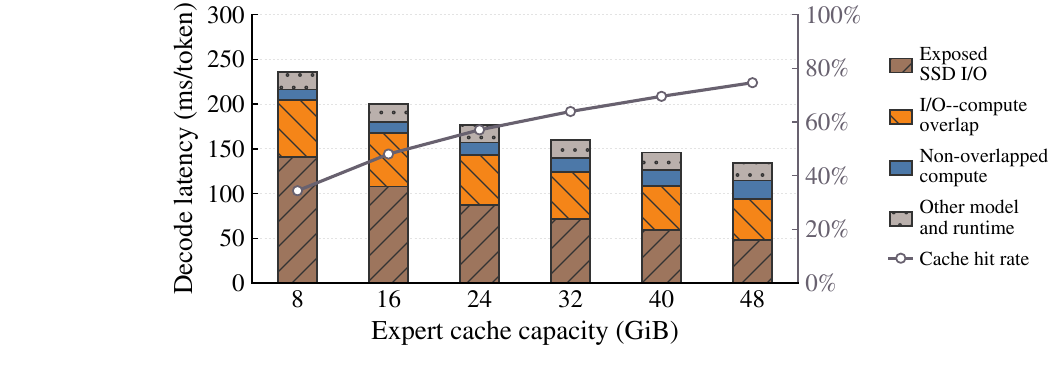}
\caption{
Cache hit rate and end-to-end decode-latency breakdown of Fixed4 on
Qwen3.5-397B-A17B. Substantial SSD latency remains
on the critical path at practical capacities.
}
\label{fig:cache-locality-and-operating-point}
\end{figure}

Figure~\ref{fig:cache-locality-and-operating-point} jointly reports
Fixed4's cache hit rate and end-to-end latency breakdown as cache
capacity increases. The rising hit rate confirms substantial temporal locality,
but practical capacities still leave substantial SSD I/O on the
critical path. With a 24~GiB expert
cache, Fixed4 achieves a 57.08\% cache hit rate but still reads
1.53~GiB from SSD per generated token. After I/O--computation overlap,
87.16~ms/token of SSD-related latency remains exposed, accounting for 49.28\% of decode
latency and limiting throughput to 5.65~tokens/s. Larger caches reduce this stall but do not eliminate it within
practical host-memory budgets.

Thus, SSD I/O remains a first-order bottleneck.
The volume of expert data drives this bandwidth demand: it determines
both how much of the expert working set fits in DRAM and how many bytes
must be transferred on each cache miss.

\section{Dynamic Expert Precision and Cross-Precision Reuse}
\label{sec:dynamic-precision}

Dynamic precision is useful only if it passes two tests. At the model
level, reducing requested expert bits must provide a favorable
quality--data tradeoff rather than assigning low precision where it
causes disproportionate error. At the system level, requests at
different precisions must preserve expert reuse throughout the storage
and execution path. This section first establishes the opportunity for
token-level precision allocation, then identifies why precision-specific
objects make this opportunity difficult to realize, and finally derives
a shared representation abstraction and its consequences for I/O and
execution.

\subsection{Token-Level Precision Reduces Expert Data Demand}
\label{sec:dynamic-precision-opportunity}

Lowering expert precision reduces data movement in two complementary
ways. First, it decreases the payload transferred when an expert is
absent from DRAM. Second, it reduces the resident footprint of that
expert, allowing a fixed-capacity cache to retain more of the expert
working set. Uniform bit-width quantization obtains both benefits, but applies
the same quality--data tradeoff to every selected expert and every
token. 

\begin{figure}[t]
  \centering

  \begin{minipage}[t]{0.45\linewidth}
    \centering
    \includegraphics[width=\linewidth]
      {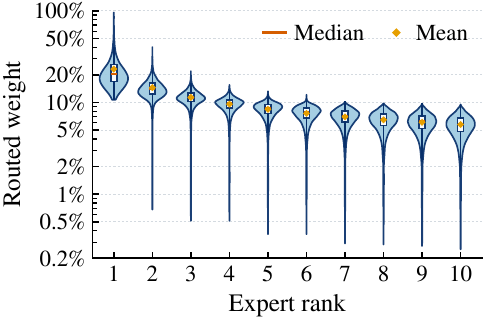}
    \parbox[t]{\linewidth}{
      \centering\footnotesize
      (a) Routing-weight distributions.
    }
  \end{minipage}%
  \hfill
  \begin{minipage}[t]{0.45\linewidth}
    \centering
    \includegraphics[width=\linewidth]
      {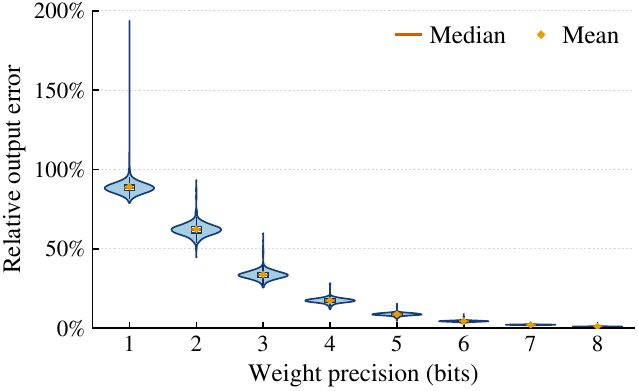}
    \parbox[t]{\linewidth}{
      \centering\footnotesize
      (b) Expert-output error distributions.
    }
  \end{minipage}

  \vspace{3pt}

  \begin{minipage}[t]{0.45\linewidth}
    \centering
    \includegraphics[width=\linewidth]
      {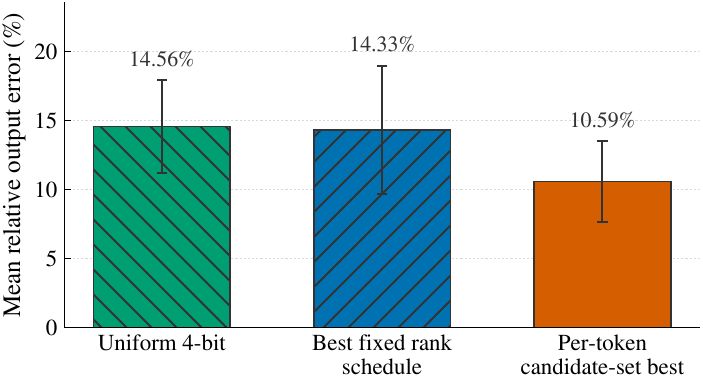}
    \parbox[t]{\linewidth}{
      \centering\footnotesize
      (c) Token-level allocation headroom.
    }
  \end{minipage}%
  \hfill
  \begin{minipage}[t]{0.45\linewidth}
    \centering
    \includegraphics[width=\linewidth]
      {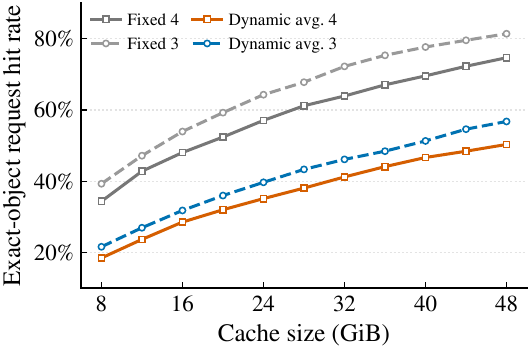}
    \parbox[t]{\linewidth}{
      \centering\footnotesize
      (d) Loss of cross-precision reuse.
    }
  \end{minipage}

  \caption{
Motivation for dynamic expert precision on Qwen3.5-397B-A17B:
(a) Routing weights vary within and across expert ranks.
(b) Experts exhibit different error reductions as precision increases.
(c) A restricted per-token candidate-set best reveals allocation
headroom under the same average-bit budget.
(d) Cache hit rates with independent precision objects. Fixed4 is from Figure~\ref{fig:cache-locality-and-operating-point},
while the other results use trace-driven simulation.
}
  \label{fig:token-level-precision}
\end{figure}

Equation~\ref{eq:routed-output} shows that the outputs of selected
experts are scaled by token-specific routing weights. The magnitudes of
these weights vary across tokens, even for experts at the same routing
rank. As shown in Figure~\ref{fig:token-level-precision}(a), the
routing-weight distributions also overlap across adjacent ranks. Rank
therefore captures only the ordering of selected experts, not the
magnitudes or relative gaps of their routing weights for the current
token.

Quantization error also changes systematically with precision. For
this characterization, we independently quantize each expert at every
evaluated bit-width using a group-wise uniform quantizer.
Figure~\ref{fig:token-level-precision}(b) shows that relative
expert-output error decreases as bit-width increases. Although the
errors are generally concentrated at each precision, low-bit
configurations exhibit a visible upper tail. Uniformly reducing
precision is therefore constrained by a subset of high-error experts,
whereas selective allocation can preserve their precision while
reducing the average expert data demand.

We use a restricted candidate-set study to characterize token-level
allocation headroom. For each token in layer~30 of
Qwen3.5-397B-A17B, we evaluate a predefined set of rank-to-bit-width
schedules under the same average 4-bit budget. The best fixed schedule
is applied to all tokens, whereas the per-token candidate-set best
selects the lowest-error schedule separately for each token using BF16
outputs as the reference. This candidate-set best is not deployable or
globally optimal; it is used only to characterize the opportunity.

As shown in Figure~\ref{fig:token-level-precision}(c), uniform 4-bit
precision produces a mean relative routed-output error of 14.56\%,
while the best fixed rank schedule achieves 14.33\%. Per-token
selection reduces the error to 10.59\%, 27.3\% lower than uniform
4-bit precision. This result does not establish
that one precision assignment is best for every model or task.
Rather, it shows that a global uniform precision leaves exploitable
token-level headroom.

Exploiting this model-level headroom in an SSD-resident runtime,
however, requires variable-precision requests to remain efficient in
the storage hierarchy. 

\subsection{The Central Obstacle: Precision-Specific Objects Break Reuse}
\label{sec:challenge-cross-precision}

A straightforward way to realize token-level dynamic precision is to
independently encode and store one expert variant for each supported
bit-width, thereby reusing conventional fixed-bit quantization formats.
Under this design, expert \(e\) has a separate representation for each supported bit-width
\(b\in\{1,\ldots,b_{\max}\}\). These precision-specific objects share
neither SSD-resident payloads nor cached expert data.

The most immediate cost is storage duplication. Storing independent 1--8-bit variants requires 36 bits per weight, \(9\times\) the payload of one 4-bit representation, even before duplicated metadata. Supporting fewer precisions reduces but does not eliminate this duplication.

More importantly, independent objects prevent cached data at one
precision from being reused by requests at another precision. Under
fixed precision, all accesses to a layer-specific expert share the
cache identity
\((\mathrm{layer},\mathrm{expert})\). Independent dynamic precision
extends this identity to
\((\mathrm{layer},\mathrm{expert},\mathrm{bitwidth})\), assigning
requests for the same expert to different precision-specific cache
objects. Two nearby requests therefore share resident data only when
their requested bit-widths also match. For example, if consecutive
tokens request the same expert at 2 and 4 bits, the resident 2-bit
object contributes no data to the separate 4-bit object, despite the
immediate reuse of the same expert.

Even if the runtime treats a resident higher-precision object as
satisfying a lower-precision request, reuse remains one-directional.
A lower-precision object still cannot be incorporated into a later
higher-precision request. The cache may therefore retain overlapping
expert information under separate identities, consuming redundant
capacity while managing and evicting each precision-specific object
independently.

Figure~\ref{fig:token-level-precision}(d) compares fixed- and
dynamic-precision caching. In the dynamic
configurations, each precision of an expert is cached as a separate object, and bit-width assignments are generated by our cache-independent
allocator. With matched average requested bit-widths and a 24~GiB cache,
Fixed4 and Fixed3 achieve hit rates of 57.08\% and 64.29\%, whereas
Dynamic Avg4 and Dynamic Avg3 achieve only 35.15\% and 39.71\%.
Thus, independent precision objects substantially reduce reuse as
requested bit-widths change.

The loss of reuse therefore arises not from the expert access sequence
or precision variation itself, but from treating each precision as a
separate stored and cached object. Avoiding this loss requires expert
state to remain shared across precision changes.

\subsection{Design Principle: Precision as Refinement Depth}
\label{sec:progressive-abstraction}

All supported bit-widths approximate the same underlying expert
weights. However, independently quantized \(b\)-bit and
\((b+1)\)-bit formats generally share no encoded data. Preserving
cross-precision reuse therefore requires a deliberately nested
representation in which different bit-widths correspond to different
refinement depths of one shared expert state.

This leads to the following design principle:

\begin{quote}
\emph{Precision should identify the refinement depth of one expert
state, not the version of an expert object.}
\end{quote}

We represent the shared state of expert \(e\) as an ordered sequence of
refinement units,

\begin{equation}
    R_e =
    \left(
        P_{e,1},
        P_{e,2},
        \ldots,
        P_{e,b_{\max}}
    \right),
\end{equation}

where \(P_{e,j}\) extends the usable precision from \(j-1\) to \(j\)
bits. A \(b\)-bit request selects refinement depth \(b\) and uses the
prefix
\(
    R_e^{(b)} =
    \left(
        P_{e,1},
        P_{e,2},
        \ldots,
        P_{e,b}
    \right).
\)

For a request to expert \(e\), let \(b_e\) denote the requested
bit-width and \(r_e\) the length of the prefix currently resident in
DRAM. Assuming that the resident refinement units form a contiguous
prefix, the additional SSD data required by the request is

\begin{equation}
    \operatorname{Fetch}(e) =
    \begin{cases}
        \varnothing,
        & b_e \leq r_e, \\[2pt]
        \{P_{e,r_e+1},\ldots,P_{e,b_e}\},
        & b_e > r_e.
    \end{cases}
\end{equation}

A resident longer prefix can directly satisfy a lower-bit request.
Conversely, a higher-bit request reuses a resident shorter prefix and
fetches only the missing suffix. Different bit-widths therefore share
one cache identity and reuse resident expert data in both request
directions, rather than selecting independent precision-specific
objects.

Realizing this abstraction requires two representation properties.
First, every supported prefix must reconstruct a valid expert
representation. Second, each additional refinement unit must be
independently addressable, so that extending a resident prefix does not
require rereading it. Prior work provides related building blocks,
including deriving multiple precisions from one representation and
organizing expert data into bit slices~\cite{park2024anyprecision,nair2025matquant}. In our SSD-resident
MoE setting, the system must preserve this nested representation across
SSD storage, DRAM caching, precision selection, and expert execution.

The shared representation resolves the cross-precision reuse problem,
but does not by itself guarantee an end-to-end reduction in data
movement. The requested bit-width must also govern the remaining token
path. First, the runtime must determine \(b_e\) before issuing expert
reads; otherwise, it cannot determine which missing refinement units
to fetch. Second, once the requested prefix is available, execution
must consume it directly; otherwise, format conversion or
full-precision materialization can recreate the data movement saved at
the SSD boundary. The next two subsections develop these consequences.

\subsection{First Consequence:
Assign Per-Expert Precision before Expert I/O}
\label{sec:challenge-preio}

Under the progressive representation, the precision assigned to an
expert determines which prefix of its representation must be
available. Precision assignment must therefore occur at a strict
boundary: after routing identifies the selected experts and their
routing weights, but before the runtime issues the corresponding SSD
reads.

Ideally, the system would choose the assignment that minimizes the
current token's routed-output deviation from BF16. Let \(e_i\) and
\(g_i(x)\) denote the \(i\)-th selected expert and its routing weight,
respectively, and let \(b_i\) denote its assigned bit-width. For an
assignment \(\mathbf{b}\), this deviation is

\begin{equation}
\label{eq:exact-current-token-error}
    \mathcal{E}(\mathbf{b};x)
    =
    \left\|
        \sum_{i=1}^{K}
        g_i(x)
        \left(
            E_{e_i}^{(b_i)}(x)
            -
            E_{e_i}^{\mathrm{BF16}}(x)
        \right)
    \right\|_2.
\end{equation}

Evaluating this quantity requires loading and executing the
corresponding expert representations. Using it to decide which
representations to load therefore creates a circular dependency: the
system would have to evaluate candidate precisions before deciding
which precision data to fetch. Enumerating candidate assignments is also impractical: with ten
selected experts and eight supported bit-widths, the unconstrained
search space contains \(8^{10}\) assignments per layer and token,
before enforcing the average-bit constraint.

Simple pre-I/O policies meet the timing requirement but discard useful information. A fixed
rank-to-precision schedule ignores current routing-weight magnitudes,
whereas a routing-only policy ignores expert- and bit-width-specific
quantization sensitivity. A practical allocator must therefore combine
both signals before expert loading, without retraining the model,
training an auxiliary selector, or performing expensive online search.

Current routing weights provide token-specific importance, while
offline expert-specific error profiles estimate the quality benefit of
each additional refinement. Independent of the current cache state,
the allocator uses these signals to select per-expert target bit-widths
\(b_e\) that minimize estimated quality loss under an average-bit
budget. The cache then maps \((b_e,r_e)\) to either prefix reuse or
missing-suffix reads; cache state thus affects the realization of the
quality--precision decision, not the decision itself.

Knowing what to read, however, addresses only the SSD-to-DRAM part of
the path. The benefit can still disappear after the requested prefix
arrives if execution first expands it into a dequantized or
full-precision expert representation.

\subsection{Second Consequence:
Execute the Selected Prefix Directly}
\label{sec:challenge-compact-execution}

Selecting the requested bit-width \(b_e\) before expert I/O avoids
unnecessary SSD reads, but reduces end-to-end data movement only if
the loaded prefix remains the
execution representation. A materialize-then-GEMV path breaks this continuity: it reads the compact prefix, writes a complete
BF16 expert matrix to DRAM, and reads that matrix again during GEMV.
At low requested bit widths, these BF16 writes and rereads can exceed the traffic of the compact input by a large margin.

GPU execution does not automatically avoid this problem. The requested
expert payload must first be staged from DRAM to device memory. On our
batch-size-one platform, completing one expert invocation directly on
the CPU takes 0.095--0.096~ms, while transferring the corresponding
low-bit payload to the GPU takes 0.114--0.121~ms from a persistent
pinned buffer and 0.31--0.32~ms with pageable blocking transfers,
excluding the subsequent GPU computation. Per-request GPU staging
therefore offers no latency advantage in this setting.

Fused low-bit execution is an established technique for avoiding
full-precision materialization. A progressively refinable
representation imposes an additional requirement: one execution path
must directly consume token-selected 1--8-bit prefixes from a cache
whose refinement units need not be physically contiguous. The kernel
must access the requested refinement units, reconstruct weight values,
and immediately consume them in GEMV.

These components form a dependency chain. The nested representation
gives different precisions one shared cache identity; pre-I/O
assignment determines the requested depth; incremental loading extends
the resident prefix; and direct execution consumes it without another
format transition. Removing any link either breaks cross-precision
reuse or reintroduces unnecessary data movement. The next section
shows how PagedMoE implements this chain.

\section{PagedMoE Overview}
\label{sec:pagedmoe-overview}

PagedMoE implements the dependency chain established in
Section~\ref{sec:dynamic-precision} through one shared progressive
expert state. For each expert \(e\), the representation \(R_e\) is
preserved across SSD storage, DRAM caching, and CPU execution. After
routing, the allocator selects a requested bit-width \(b_e\), which
specifies the required prefix length for the current token, while the
cache maintains the resident prefix length \(r_e\) across tokens. The
runtime compares \(b_e\) with \(r_e\) to determine whether the request
can reuse the cached prefix directly or requires additional refinement
units. Once the requested prefix \(R_e^{(b_e)}\) of an expert is available, the
CPU executes it without constructing a separate precision-specific
expert representation.

\begin{figure}[t]
  \centering
  \includegraphics[width=\linewidth]{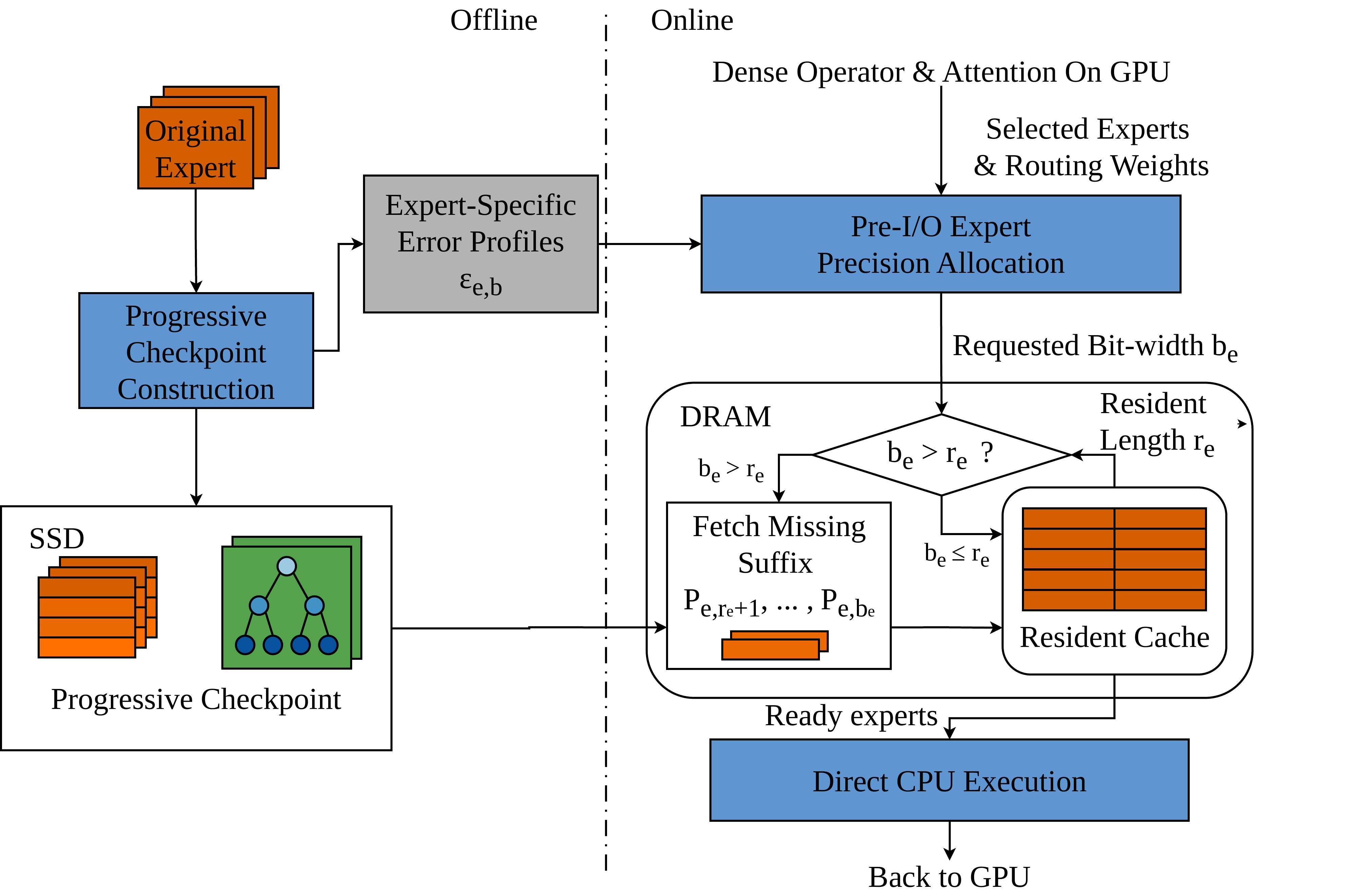}
  \caption{
  PagedMoE's shared-state workflow. Offline processing constructs one
  progressive representation and an error profile for each expert.
  Online, routing and error profiles determine the requested bit-width
  \(b_e\); the cache reuses the resident prefix or fetches only its
  missing suffix according to \(r_e\); and the CPU directly executes 
  each requested prefix once it is ready.
  }
  \label{fig:pagedmoe-overview}
\end{figure}

Figure~\ref{fig:pagedmoe-overview} summarizes this flow from offline
artifacts to online execution. 
Before deployment, PagedMoE constructs two artifacts. The first is a
progressive checkpoint that stores each expert as bitplanes
\(R_e=(P_{e,1},\ldots,P_{e,b_{\max}})\) with prefix-compatible tree
codebooks. Each \(P_{e,j}\) is an independently addressable refinement
unit, and the codebooks make every prefix \(R_e^{(b)}\) a valid
\(b\)-bit reconstruction. Thus, all bit-widths share one encoded expert
rather than independently quantized copies
(Section~\ref{sec:progressive-representation}). The second consists of
expert-specific error profiles measured from random calibration inputs;
they estimate approximation error at each supported bit-width and guide
online allocation.

During decoding, the GPU executes dense operators, attention, routing,
and other always-activated components. For every selected expert, the
allocator combines its current routing weight with its offline error
profile and assigns a target bit-width \(b_e\) under an average
requested-bit budget. This decision is made after routing but before
any expert read, so the selected precision can directly control the
data requested from SSD. Section~\ref{sec:precision-allocation}
presents the allocation objective and algorithm.

The runtime then compares \(b_e\) with the cached resident length
\(r_e\). If \(b_e\leq r_e\), the requested prefix is already available
and the expert becomes ready immediately. Otherwise, the runtime
loads only the missing bitplanes
\(P_{e,r_e+1},\ldots,P_{e,b_e}\), preserving and extending the prefix
loaded by previous tokens. Cache-ready experts can execute while SSD
reads for other experts remain in flight.

Once the requested prefix is complete, the CPU kernel consumes
the same cached prefix directly. It reconstructs weight values from
the requested bitplanes and immediately uses them in GEMV, without
repacking the expert into a persistent compute representation or
materializing a complete BF16 weight matrix. The resulting expert
activations are combined and then return to the GPU for the remaining model operators.

The representation, allocator, and runtime play distinct roles in one
progressive-state workflow. The representation defines
what can be shared, the allocator determines how much precision the
current token requires, and the runtime reuses or incrementally
extends the resident state before executing the requested prefix. No
stage turns a precision change into a separate expert object.
Sections~\ref{sec:progressive-representation}--
\ref{sec:runtime} present these stages in detail.

\section{A Single-Copy Expert Representation for Cross-Precision Reuse}
\label{sec:progressive-representation}

\subsection{Design Rationale}

To enable cross-precision reuse, the progressively refinable representation
\(R_e\) must satisfy two properties. First, the first \(b\) bits of each
element must define a valid \(b\)-bit reconstruction.
Second, the refinement from \(b\) to \(b+1\) bits must be independently
addressable, allowing the runtime to fetch it without rereading the resident
prefix.

A standard fixed-\(k\)-bit codebook defines reconstructions only for complete
indices. PagedMoE observes that a \textbf{binary prefix tree} can serve as a
precision hierarchy: each non-root node represents an index prefix and is
assigned a reconstruction value, while truncating an index moves to an
ancestor on the same path. Prefix length therefore represents usable
precision.

A tree alone, however, does not make successive index bits independently
addressable in storage. PagedMoE therefore pairs it with an expert-level
bitplane layout, where each bitplane \(P_{e,j}\) stores the \(j\)-th index bit
of every element and serves as one independently loadable refinement unit.
Together, they make a precision change an extension of the same encoded
expert rather than a switch to a separate precision object.

\subsection{Tree-Structured Codebook}
\label{sec:tree-quantization}

\paragraph{Maximum-Depth Codebook Quantization}
PagedMoE partitions each routed-expert tensor into quantization blocks.
For a block \(W=\{w_1,\ldots,w_N\}\), it learns a maximum-depth
\(b_{\max}\)-bit codebook \(\mathcal{C}^{(b_{\max})}\) containing
\(2^{b_{\max}}\) reconstruction values and assigns each weight an index
\(z_i\in\{0,\ldots,2^{b_{\max}}-1\}\). Weight \(w_i\) is reconstructed as
\(C_{z_i}\), and the codebook and index assignments are learned by minimizing

\[
\min_{\mathcal{C}^{(b_{\max})},\mathbf{z}}
\sum_{i=1}^{N}
\left(w_i-C_{z_i}\right)^2.
\]

Because the weights are scalar, this is a one-dimensional \(K\)-means
problem. PagedMoE uses an optimized offline solver to make codebook
construction practical at checkpoint scale.

\paragraph{Tree-Structured Prefix Reconstruction}
PagedMoE transforms the learned maximum-depth codebook into a binary prefix
tree. It first sorts the \(2^{b_{\max}}\) codewords by reconstruction value and
remaps their indices accordingly. The sorted codewords are placed at the leaves from left to right and merged
bottom up. Each pair of adjacent nodes whose indices differ only in the
last bit shares a parent obtained by removing that bit. As the codewords were sorted, nearby
reconstruction values are merged before more distant ones.

Each internal node represents the index prefix shared by all leaves in its
subtree. PagedMoE assigns the node the frequency-weighted average of its
descendant leaf values, which minimizes the squared
reconstruction error for the fixed leaf assignment and tree structure.
\begin{figure}[t]
\centering

\begin{minipage}[b]{0.37\linewidth}
  \centering
  \includegraphics[width=\linewidth]{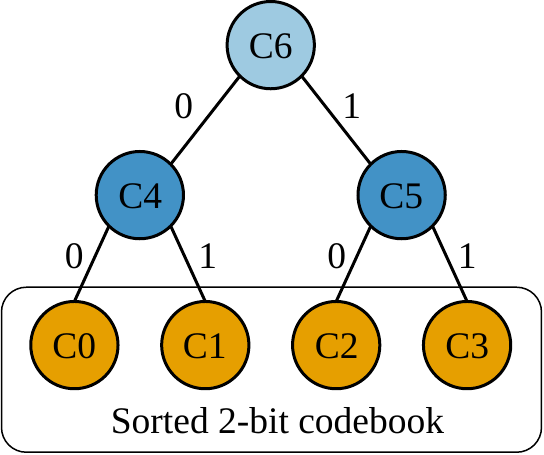}
  \par
  {\footnotesize (a)}
\end{minipage}
\hfill
\begin{minipage}[b]{0.59\linewidth}
  \centering
  \includegraphics[width=\linewidth]{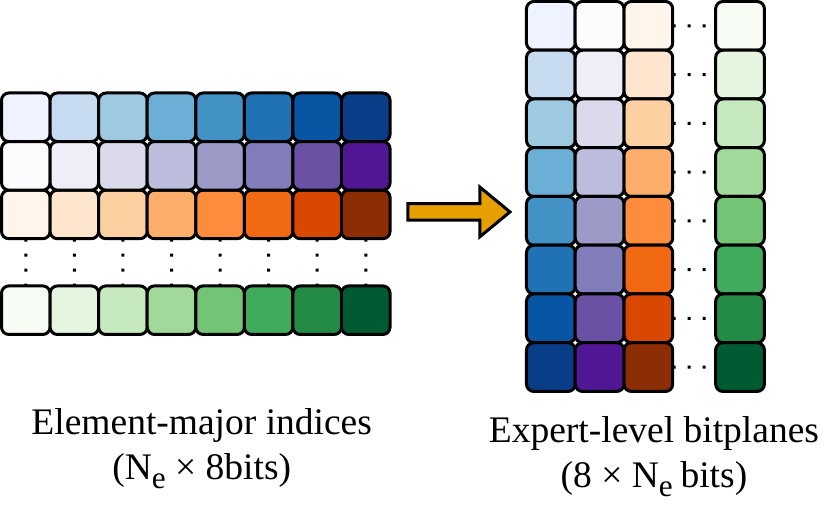}
  \par
  {\footnotesize (b)}
\end{minipage}

\caption{
PagedMoE's progressive expert representation:
(a) a tree codebook assigns a valid reconstruction to every supported
index prefix; and
(b) expert-level bitplanes make each additional index bit an independently
addressable refinement unit.
}
\label{fig:progressive-representation}
\end{figure}
Figure~\ref{fig:progressive-representation}(a) illustrates the construction for a
2-bit codebook. After sorting, \(C_0,C_1,C_2,C_3\) are assigned indices
\(00,01,10,11\), respectively. The leaves \(C_0\) and \(C_1\) share the
1-bit prefix \(0\), and their parent \(C_4\) provides the corresponding
1-bit reconstruction:

\[
C_4
=
\frac{f_0C_0+f_1C_1}
     {f_0+f_1},
\]

where \(f_0\) and \(f_1\) are their assignment frequencies. The remaining
internal nodes are constructed in the same manner.

A weight assigned the maximum-depth index \(01\), for example, uses \(C_1\)
at 2-bit precision and its parent \(C_4\) at 1-bit precision. In general, a
\(b\)-bit request follows the first \(b\) bits of the stored index and uses
the node at depth \(b\). All supported precisions therefore derive from one
maximum-depth index rather than from independently quantized copies.

The lower-precision codebooks are derived from the maximum-depth assignment
rather than optimized independently. We evaluate the resulting quantization
quality in Section~\ref{sec:evaluation}.

\subsection{Bitplane Layout for Incremental Refinement}
\label{sec:bitplane-layout}

Prefix-valid reconstruction alone does not make precision refinement
efficient. In an element-major layout, the \(b_{\max}\) bits of each weight index are packed
together. Refining from \(x\) to \(x+1\) bits requires the \((x+1)\)-th bit
of every index, but fetching these interleaved bits also reads unnecessary
or already-resident data, causing SSD read amplification and additional
cache-update work.

PagedMoE therefore transposes the maximum-depth indices into expert-level
bitplanes, as shown in Figure~\ref{fig:progressive-representation}(b). For expert \(e\),
\(P_{e,j}\) concatenates the \(j\)-th index bit of every quantized weight
across all blocks and is stored contiguously as one independently addressable
refinement unit.

If \(P_{e,1},\ldots,P_{e,x}\) are resident, refining the expert to
\(x+1\) bits requires fetching only \(P_{e,x+1}\), without rereading or
reorganizing the prefix. Higher-precision requests therefore extend the same
cached state, while lower-precision requests reuse its prefix.

\subsection{Storage and Metadata Overhead}

PagedMoE partitions each routed-expert weight tensor into quantization blocks of
\(N\) elements and constructs one tree-structured codebook for each block. Each encoded expert therefore stores multiple block-level codebooks
and \(b_{\max}\) expert-level bitplanes containing the transposed index bits.

With \(b_{\max}=8\), each block stores 256 leaf and 254 non-root internal-node
values, totaling 1020 bytes in BF16. For \(N\geq2^{17}\) in our setting, this metadata is
less than \(0.8\%\) of the 8-bit index payload. Although the payload is twice
that of a fixed-4-bit representation, one copy exposes all 1--8-bit prefixes
instead of requiring separate per-precision copies.

This representation exposes each expert as an ordered sequence of
independently loadable refinement bitplanes. The next section determines, before I/O,
how many of these bitplanes each selected expert should request.

\section{Pre-I/O Expert Precision Allocation}
\label{sec:precision-allocation}

The representation in Section~\ref{sec:progressive-representation} exposes
each expert as an ordered sequence of refinement bitplanes. After routing,
PagedMoE must determine the target bit-width \(b_i\) of each selected expert
before issuing its SSD reads. This target determines how much of the expert
representation must be available, while the runtime later determines which
of the required bitplanes are already resident.

At this boundary, the current routing weights are available, but the
current-token quantization error at each candidate bit-width is not:
evaluating it would require loading and executing the corresponding expert
prefix. PagedMoE therefore combines current routing weights with offline
expert-specific error profiles to assign target bit-widths under an average
requested-bit budget.

For each token, the router selects
\(\mathcal{S}=\{e_1,\ldots,e_K\}\) with normalized routing weights
\(g_1,\ldots,g_K\). The allocator assigns each selected expert a bit-width
\(b_i\in\mathcal{B}=\{1,\ldots,b_{\max}\}\) subject to

\[
\sum_{i=1}^{K} b_i \leq cK,
\]

where \(c\) is the target average bit-width and is chosen such that
\(cK\) is an integer. This budget constrains requested precision rather
than cache-miss bytes. Given the resulting bit-widths, the
runtime determines the actual SSD reads from the current cache state.

\subsection{Offline Expert Error Profiles}
\label{sec:expert-error-profiles}

PagedMoE profiles each expert's quantization sensitivity using a small
set of dataset-independent random calibration inputs
\(\mathcal{Z}=\{z_1,\ldots,z_n\}\). The same calibration set is used
for all experts. For each expert \(e\) and supported bit-width \(b\),
PagedMoE defines the offline sensitivity score as

\[
\epsilon_{e,b}
=
\left(
\frac{1}{|\mathcal{Z}|}
\sum_{z\in\mathcal{Z}}
\frac{
\left\|
E_e^{(b)}(z)-E_e^{\mathrm{BF16}}(z)
\right\|_2
}{
\|z\|_2
}
\right)^2.
\]

Here, \(E_e^{(b)}\) and \(E_e^{\mathrm{BF16}}\) denote the
\(b\)-bit and reference BF16 experts, respectively. Input normalization
makes the scores comparable across calibration inputs. 

The resulting profile
\(
(\epsilon_{e,1},\ldots,\epsilon_{e,b_{\max}})
\)
characterizes expert \(e\)'s approximation sensitivity across
different prefix depths. Because it depends only on the encoded expert
and the fixed random calibration inputs, PagedMoE computes it once
offline.

\subsection{Routing-Weighted Error Surrogate}
\label{sec:error-surrogate}

The offline profiles capture expert-specific sensitivity but not an expert's
importance to the current token. PagedMoE incorporates this token-specific
information through the current routing weights.

As discussed in Section~\ref{sec:challenge-preio}, the exact current-token
objective in Equation~\ref{eq:exact-current-token-error} cannot be evaluated
before expert loading. Squaring that objective and normalizing it by
\(\|x\|_2^2\), which is constant for the current token, yields per-expert
error terms scaled by \(g_i^2\) and cross-expert inner products.
Motivated by this decomposition, PagedMoE constructs a separable surrogate
by omitting the unavailable cross-expert terms and using the offline profile
as an expert- and precision-specific sensitivity score:

\[
\widehat{\operatorname{Err}}^2(\mathbf{b})
=
\sum_{i=1}^{K}
g_i^2 \epsilon_{e_i,b_i}.
\]

The factor \(g_i^2\) reflects routing-weight scaling in the squared-error
objective, while \(\epsilon_{e_i,b_i}\) captures its profiled
sensitivity at the selected bit-width. Both quantities are available before
SSD I/O.

PagedMoE selects the target bit-widths by solving

\[
\begin{aligned}
\min_{\mathbf{b}}\quad
 \sum_{i=1}^{K} g_i^2\epsilon_{e_i,b_i} \quad \quad
\text{s.t.}
 \sum_{i=1}^{K} b_i \leq cK, \quad
 b_i\in\mathcal{B}.
\end{aligned}
\]

\subsection{Marginal-Gain Greedy Allocation}
\label{sec:marginal-gain-allocation}

PagedMoE initially assigns the minimum supported 1-bit prefix to every
selected expert. The reduction in the surrogate objective obtained by
refining expert \(e_i\) from \(b-1\) to \(b\) bits is

\[
\Delta_{i,b}
=
g_i^2
\left(
\epsilon_{e_i,b-1}
-
\epsilon_{e_i,b}
\right), \qquad 2\leq b\leq b_{\max}.
\]

A large \(\Delta_{i,b}\) indicates that the expert has a large routing
weight, that the additional bit substantially reduces its profiled error, or
both.

Successive refinements generally exhibit diminishing returns. Across all
profiled experts in Qwen3.5-397B-A17B, \(99.45\%\) of the profiles
already have nonnegative, non-increasing successive error reductions as the
bit-width increases. For the remaining profiles, PagedMoE first clips
negative successive error reductions to zero and then applies equal-weight
isotonic regression using the pool-adjacent-violators algorithm
(PAVA)~\cite{best1990active} to enforce a non-increasing sequence. The
allocator uses these corrected reductions to compute the marginal gains.
Because \(g_i^2\) is constant across all refinements of expert \(e_i\), the
resulting marginal gains satisfy

\[
\Delta_{i,2}
\geq
\Delta_{i,3}
\geq
\cdots
\geq
\Delta_{i,b_{\max}}
\geq 0. 
\]

\begin{figure}[t]
\centering
\includegraphics[width=\linewidth]{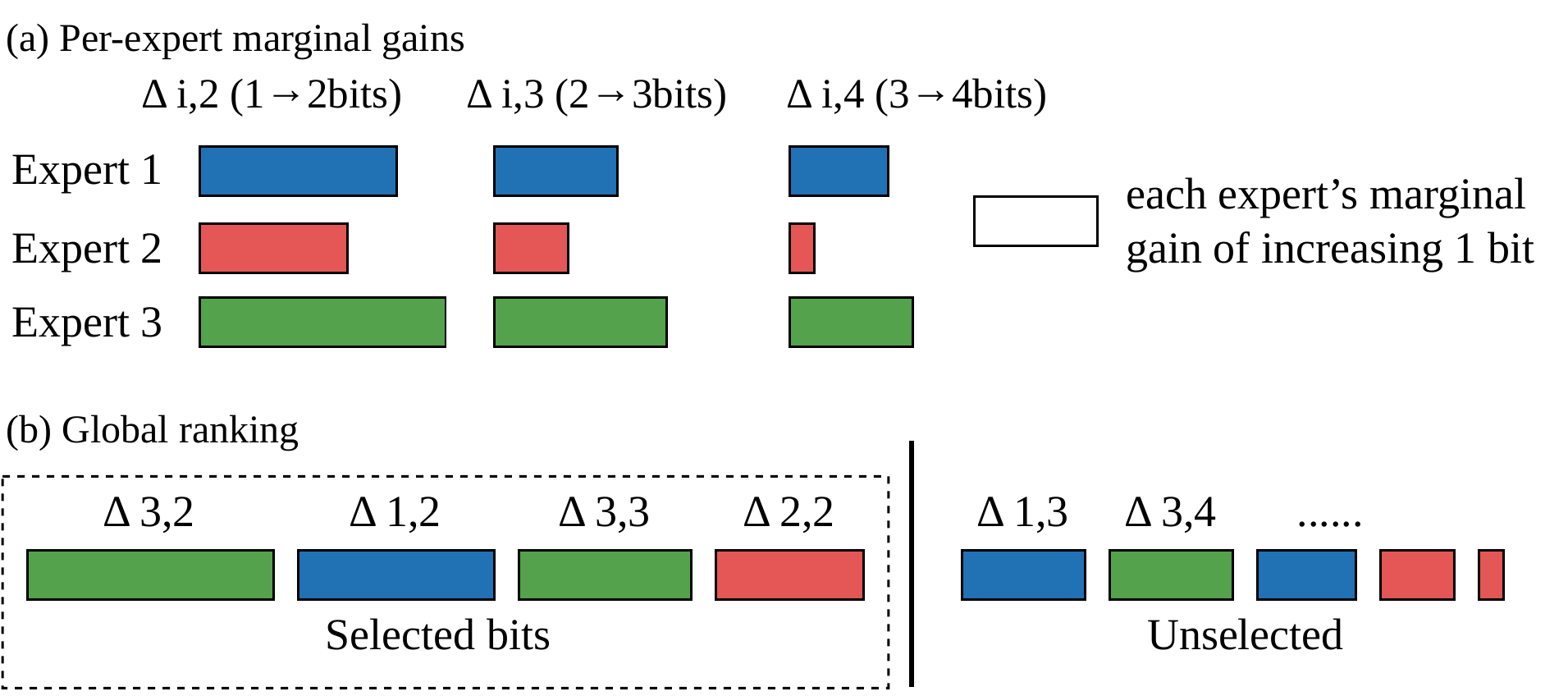}
\caption{
Marginal-gain precision allocation.
Each selected expert contributes a sequence of routing-weighted refinement
gains that decreases with precision. PagedMoE globally ranks all candidate
refinements and selects the largest \((c-1)K\) gains under the bit budget.
The diminishing-return property ensures that the selected refinements for
each expert form a valid precision prefix.
}
\label{fig:marginal-gain-allocation}
\end{figure}

As illustrated in Figure~\ref{fig:marginal-gain-allocation}, PagedMoE
computes all \(K(b_{\max}-1)\) candidate gains, sorts them globally, and
selects the largest \((c-1)K\) refinements. The assigned bit-width of expert
\(e_i\) is one plus the number of its selected refinements.

Because the corrected gains are non-increasing within each expert, selecting
a refinement also selects all of its predecessors; ties are resolved in
favor of lower-bit refinements. The selected refinements therefore form a
valid prefix for every expert. Equivalently, after this correction, the problem is separable
discrete-concave resource allocation, for which global marginal-gain
selection is optimal~\cite{federgruen1986greedy,groenevelt1991resource}.

Computing and sorting the gains costs
\(O(Kb_{\max}\log(Kb_{\max}))\) time and \(O(Kb_{\max})\) temporary space;
with \(K=10\) and \(b_{\max}=8\), each layer has only 70 candidates.

The allocator therefore produces one target bit-width \(b_i\) for each
selected expert before any expert read is issued. The runtime in the next
section compares these targets with the resident prefixes and fetches only
the missing bitplanes.

\section{Progressive-State Runtime}
\label{sec:runtime}

The allocator in Section~\ref{sec:precision-allocation} produces a
target depth \(b_e\) for every selected expert before expert I/O.
The runtime maintains the resident prefix length \(r_e\) across tokens
and realizes each request by extending or reusing that state. If
\(b_e\leq r_e\), the cached prefix can serve the request immediately;
otherwise, the runtime loads only the missing refinement bitplanes.
Once the complete requested prefix is available, the expert becomes
ready and is executed directly from the same cached representation.

PagedMoE is integrated with SGLang~\cite{sglang} for batch-size-one
local inference.
Dense and always-activated operators execute on the GPU, while routed
experts remain SSD-resident and execute on the CPU. The Rust runtime
manages bitplane residency, asynchronous SSD requests, and task
scheduling, while performance-critical expert kernels are implemented
in C/C++ using AVX-512.

\subsection{Prefix-Preserving Cache State}
\label{sec:bitplane-cache}

PagedMoE manages routed-expert weights at the granularity of expert-level bitplanes. A cache entry \((e,j)\) contains the \(j\)-th index bit of every quantized weight in expert \(e\) and serves as a unit of DRAM allocation, SSD I/O, and replacement. An expert's tree codebook is loaded and evicted together with its first bitplane, but stored separately in DRAM. Consequently, every executable resident prefix has the codebook required to reconstruct its weights.

For each expert, \(r_e\) is the number of consecutive bitplanes resident from \(P_{e,1}\) onward. PagedMoE enforces a \emph{prefix-residency invariant}: a bitplane is usable only if all earlier bitplanes are also available.

PagedMoE manages these bitplanes using an LRU cache. When a \(b_e\)-bit prefix is accessed, the runtime refreshes its entries from \(P_{e,b_e}\) down to \(P_{e,1}\). This makes \(P_{e,1}\) the most recent and \(P_{e,b_e}\) the least recent among the accessed bitplanes. Higher-depth bitplanes are therefore evicted before lower-depth ones, shortening the resident prefix without leaving an internal hole.

The resident bitplanes of an expert need not occupy a contiguous physical region in DRAM. PagedMoE therefore neither reserves a maximum-depth physical buffer for every resident expert nor assembles its bitplanes into a persistent packed copy. When an expert task begins execution, the cache keeps its requested bitplanes valid until the kernel completes.

\subsection{Missing-Suffix Loading and Pipelined Scheduling}
\label{sec:runtime-loading}

\begin{figure}[t]
\centering
\includegraphics[width=0.82\linewidth]{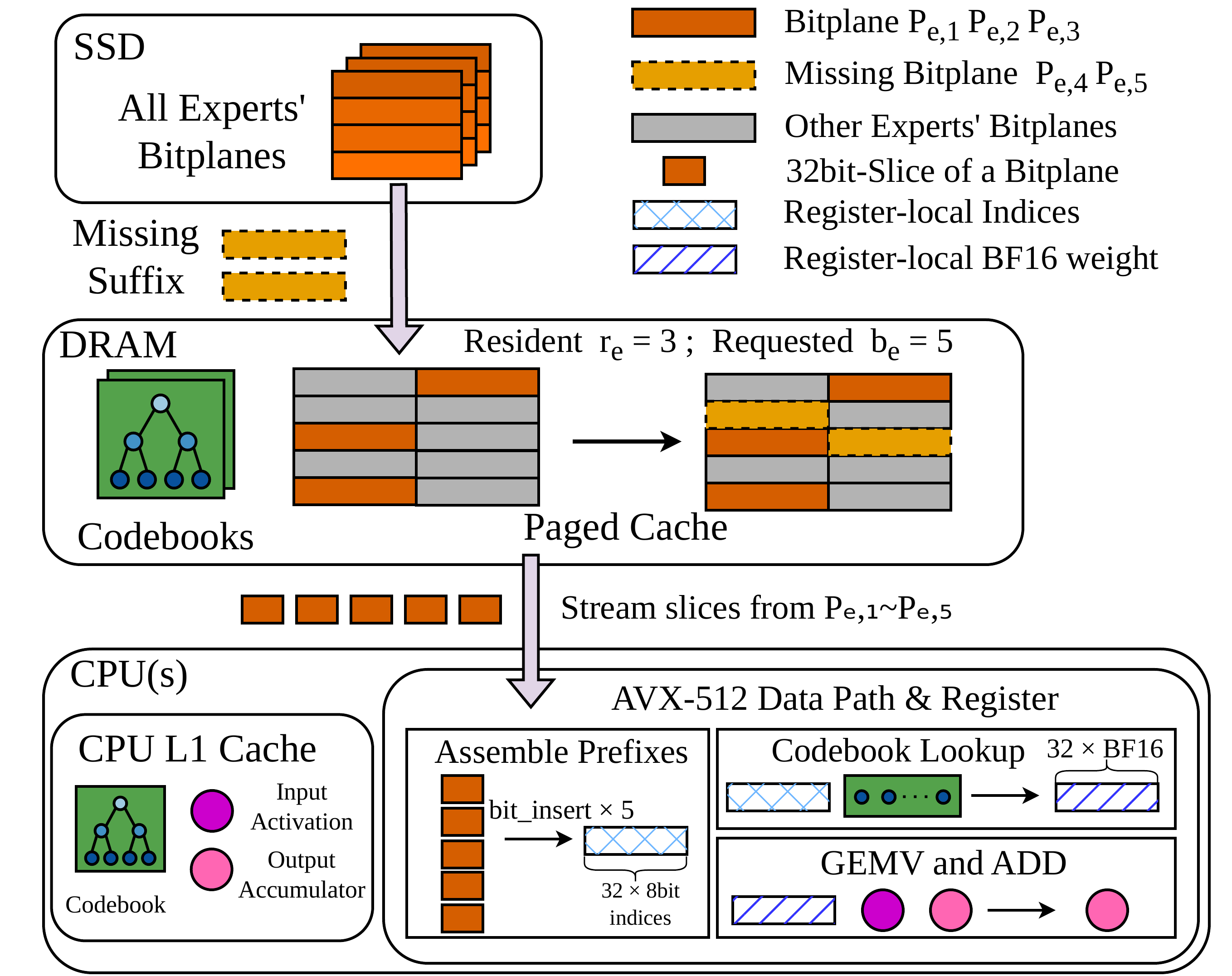}
\caption{PagedMoE's bitplane cache and direct-execution path. For \(r_e=3\)
and \(b_e=5\), it fetches only \(P_{e,4}\) and \(P_{e,5}\); the CPU
reconstructs weights on the fly from cached bitplanes and immediately
consumes them in GEMV without materializing a BF16 matrix.}
\label{fig:bitplane-cache}
\end{figure}

For each selected expert, the runtime compares its target depth \(b_e\) with its resident length \(r_e\). If \(b_e\leq r_e\), the expert is ready immediately. Otherwise, the runtime asynchronously reads only \(P_{e,r_e+1},\ldots,P_{e,b_e}\), reusing the complete resident prefix. Although these reads may complete out of order, the expert becomes ready only after its codebook and all bitplanes required by \(b_e\) are available. Figure~\ref{fig:bitplane-cache} illustrates this process for \(r_e=3\) and \(b_e=5\), where only \(P_{e,4}\) and \(P_{e,5}\) are fetched.

Ready expert tasks are dispatched to the CPU immediately while SSD reads for other experts continue. This pipelines expert computation with the remaining I/O, avoiding a barrier that waits for all selected experts in the layer to finish loading. The schedule does not change the amount of requested data; it only overlaps part of the residual SSD latency after dynamic precision and cross-precision reuse have reduced the traffic.

\paragraph{Explicit runtime management.}
An mmap-backed implementation can also expose noncontiguous physical pages through a contiguous virtual address range. Physical contiguity is therefore not the reason for explicit paging. PagedMoE instead manages bitplanes explicitly so that prefix residency, missing refinements, in-flight I/O, replacement, and task readiness remain visible to the scheduler. Section~\ref{sec:eval-explicit-runtime} compares this design with an mmap-backed runtime under total-memory accounting.

\subsection{Direct Execution of the Requested Prefix}
\label{sec:direct-execution}

Once an expert becomes ready, the CPU kernel directly reads its first
\(b_e\) cached bitplanes, which may reside in different DRAM cache
blocks. The kernel streams corresponding slices from these bitplanes
without merging them into a persistent element-major representation or
reconstructing a complete BF16 expert matrix.

Figure~\ref{fig:bitplane-cache} illustrates one SIMD step. For the next
32 weights, the kernel reads a 32-bit slice from each of the \(b_e\)
bitplanes. Each slice contributes one bit to 32 weight indices, and
AVX-512 bit-insertion operations assemble them into 32 register-local
8-bit indices.

The reconstruction values at the requested codebook depth are loaded
into SIMD registers for lookup. When \(b_e\leq5\), all
\(2^{b_e}\) BF16 codebook entries fit in one 512-bit SIMD register, so the
32 indices can be resolved with a single vector lookup. At higher
precisions, the codebook entries span multiple registers and the
kernel combines the results of multiple lookups. The resulting 32 BF16
weights remain in a SIMD register and are immediately consumed by BF16
dot-product instructions, with the results accumulated in FP32. The kernel also unrolls four weight-matrix rows to expose
instruction-level parallelism across independent output accumulations.

The expert computation is partitioned into blocks and distributed
across CPU workers. Each worker processes one block using the
corresponding segment of the input activation, while multiple workers
execute independent blocks in parallel. Reconstructed weights are
never written back as an expert-scale intermediate matrix.

The same bitplane-native execution path supports target depths from
1 to 8 bits. Numerical reconstruction is fused with GEMV rather than
performed as a separate full-expert materialization stage. The
progressively refinable representation therefore remains compact
across SSD storage, DRAM caching, and CPU execution.
Section~\ref{sec:eval-compact-execution} evaluates the benefit of this
direct execution path.

\section{Evaluation}
\label{sec:evaluation}

We evaluate PagedMoE in three steps. First, we report end-to-end
quality and decode throughput. Second, we separate the benefit of
requesting fewer expert bits from the additional benefit of sharing
cached state across precisions. Third, we evaluate prefix-compatible
quantization, pre-I/O allocation, and direct compact execution, and then provide a supporting complete-runtime comparison with mmap.

\subsection{Experimental Setup}
\label{sec:eval-setup}

\noindent\textbf{Hardware.}
We evaluate on a local PC with an AMD Ryzen 9 9950X CPU
(16 cores and 32 threads), an NVIDIA RTX 5090 GPU (32~GiB),
64~GB of dual-channel DDR5-6400 memory, and a 4~TB PCIe
5.0\(\times4\) NVMe SSD. Although the GPU provides 32~GiB of memory, PagedMoE uses it only for dense and always-activated components; routed experts remain SSD-resident and execute on the CPU.

\noindent\textbf{Models and quantization scope.}
We use Qwen3.5-397B-A17B~\cite{qwen35_397b}, which selects 10 of
512 experts per MoE layer, as the primary model, and GLM5.2~\cite{glm52},
which selects \(K=8\), to test generalization. PagedMoE quantizes only
routed-expert weights, retaining other Qwen3.5 components in BF16 and
other GLM5.2 components at FP8/BF16~\cite{glm52_fp8}. The official Qwen GPTQ-INT4 checkpoint likewise quantizes only
routed-expert weights~\cite{qwen35_397b_gptq_int4,gptq}. For GLM5.2, we also
derive a calibration-free, weight-only MXFP4 baseline~\cite{rouhani2023microscalingdataformatsdeep}.

\noindent\textbf{System configurations.}
\emph{Uniform4} always requests four bits using PagedMoE's
prefix-preserving cache; unlike the motivation study's Fixed4, it can
retain and incrementally extend shorter prefixes. \emph{Dynamic Avg4}
and \emph{Dynamic Avg2.5} use PagedMoE's allocator with average
requested-bit budgets of 4 and 2.5 bits.
\emph{Independent} replays the same requests and bit-width assignments
under an otherwise identical runtime but disables cross-precision reuse. We also evaluate \texttt{ik\_llama.cpp}~\cite{ik_llama_cpp}
with \texttt{UD-IQ4} checkpoints~\cite{qwen35_ud_iq4_xs,glm52_ud_iq4_xs} and mmap-backed page caching.
An mmap-backed variant uses the same progressive checkpoint and
precision assignments but relies on operating-system demand paging.

\noindent\textbf{PagedMoE configuration and calibration.}
PagedMoE uses blocks of \(2^{17}\) weights for Qwen3.5 and
\(3\times2^{17}\) weights for GLM5.2, encoding each routed-expert
tensor once to a maximum depth of 8 bits. Error profiles use 10
Gaussian BF16 inputs, shared across experts within
each model and independent of evaluation data.

\noindent\textbf{Memory accounting.}
Cache-budget comparisons use PagedMoE's configured cache capacity; for \texttt{ik\_llama.cpp}, we set the cgroup memory limit to the target cache size plus the empirically measured non-mmap memory usage: 1.27~GiB for Qwen3.5 and 2.36~GiB for GLM5.2. Total-memory comparisons count PagedMoE's
cache plus 5.2~GiB of measured non-cache memory, and use the actual
cgroup \texttt{MemoryMax} for mmap-backed configurations.

\noindent\textbf{Metrics and protocol.}
We report average requested bits per selected-expert invocation, byte
hit rate as the resident fraction of requested bitplane bytes, and SSD
traffic from logged direct-I/O requests that bypass the operating-system
page cache. Each batch-size-one run starts with an empty expert cache;
after prefill, which may warm the cache, we measure 500 decode tokens%
. Quality is evaluated with the
EleutherAI Language Model Evaluation Harness~\cite{lm_eval_harness}
on benchmarks~\cite{gsm8k,openbookqa,arc,winogrande,gpqa,wikitext2}.

\subsection{Overall Quality and End-to-End Performance}
\label{sec:eval-overall}

\begin{table}[t]
\centering
\small
\setlength{\tabcolsep}{3pt}
\caption{End-to-end quality on Qwen3.5-397B-A17B and GLM5.2. Higher is better except
for perplexity.}
\label{tab:quality}
\begin{tabular}{lcccc}
\toprule
\multicolumn{5}{l}{\textbf{Qwen3.5-397B-A17B}} \\
\textbf{Benchmark} & \textbf{BF16} & \textbf{GPTQ-INT4} &
\textbf{Uniform4} & \textbf{Dyn Avg2.5} \\
\midrule
GSM8K-CoT      & 97.50\% & 97.12\% & 97.04\% & 96.89\% \\
OpenBookQA     & 47.40\% & 46.60\% & 46.60\% & 47.80\% \\
ARC-Challenge  & 68.43\% & 67.58\% & 68.69\% & 67.75\% \\
WinoGrande     & 78.69\% & 77.35\% & 79.08\% & 78.37\% \\
GPQA Diamond   & 54.55\% & 53.03\% & 51.01\% & 52.53\% \\
WikiText-2 PPL & 3.171   & 3.201   & 3.223   & 3.240   \\
\midrule
\multicolumn{5}{l}{\textbf{GLM5.2}} \\
\textbf{Benchmark} & \textbf{FP8-official} & \textbf{MXFP4} &
\textbf{Uniform4} & \textbf{Dyn Avg2.5} \\
\midrule
GSM8K-CoT      & 92.57\% & 93.40\% & 93.63\% & 90.52\% \\
OpenBookQA     & 48.20\% & 48.40\% & 47.40\% & 48.40\% \\
ARC-Challenge  & 70.05\% & 69.11\% & 70.14\% & 69.62\% \\
WinoGrande     & 75.22\% & 74.74\% & 76.56\% & 74.82\% \\
GPQA Diamond   & 46.97\% & 43.43\% & 49.49\% & 45.45\% \\
\bottomrule
\end{tabular}
\end{table}

\paragraph{End-to-end quality.}
Table~\ref{tab:quality} reports results on both models. On
Qwen3.5-397B-A17B, Uniform4 achieves quality comparable to the
official GPTQ-INT4 checkpoint, showing that the prefix-compatible
representation provides competitive four-bit quality. Dynamic Avg2.5
reduces requested routed-expert bits by 37.5\% relative to Uniform4
while remaining close to GPTQ-INT4 across all evaluated tasks. These
results show that token-level allocation can substantially reduce
expert-data demand by distributing precision nonuniformly across
selected experts.

On GLM5.2, Uniform4 similarly remains comparable to MXFP4, supporting
the quality of the prefix-compatible representation on a second model.
Dynamic Avg2.5 matches or exceeds MXFP4 on all benchmarks except GSM8K-CoT. Thus, the representation and allocation trends generalize,
although the quality cost of a 2.5-bit average budget remains
model- and task-dependent.

\begin{figure}[t]
\centering
\begin{minipage}[t]{0.485\linewidth}
\centering
\includegraphics[width=\linewidth, trim=0 1mm 0 1.5mm, clip]
{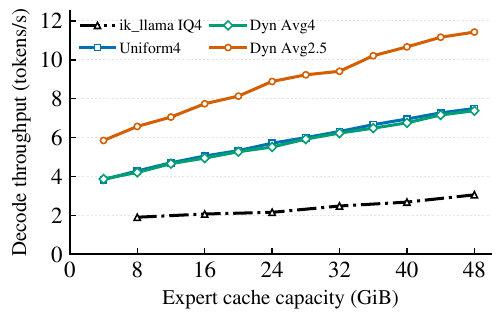}
{\small (a) Qwen3.5-397B-A17B.}
\end{minipage}%
\hfill
\begin{minipage}[t]{0.485\linewidth}
\centering
\includegraphics[width=\linewidth, trim=0 1mm 0 1.5mm, clip]
{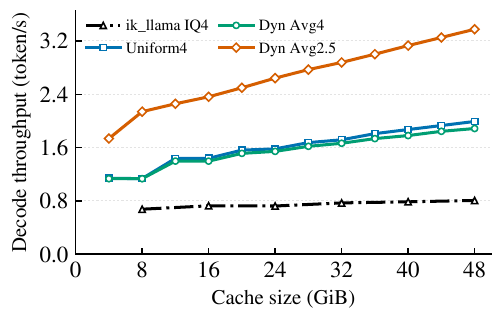}
{\small (b) GLM5.2.}
\end{minipage}
\caption{
End-to-end batch-size-one decode throughput versus configured
expert-cache capacity on (a) Qwen3.5-397B-A17B and (b) GLM5.2.
}
\Description{
Two line charts comparing end-to-end decoding throughput under
different routed-expert cache budgets on Qwen3.5-397B-A17B and
GLM5.2.
}
\label{fig:e2e-throughput}
\end{figure}

\paragraph{End-to-end throughput.}
Figure~\ref{fig:e2e-throughput} compares batch-size-one decode throughput
under routed-expert cache-budget accounting.

On Qwen3.5-397B-A17B, Dynamic Avg4 closely tracks Uniform4 at the
same average requested-bit budget. At a 24~GiB cache capacity,
Uniform4 reaches 5.71~tokens/s and Dynamic Avg4 reaches
5.52~tokens/s, showing little throughput overhead from variable
per-expert precision. Reducing the average budget to 2.5 bits
increases throughput to 8.89~tokens/s, \(1.56\times\) that of
Uniform4. This combined benefit is decomposed into fewer requested
bits and cross-precision reuse in
Section~\ref{sec:eval-cross-precision}.

For reference, \texttt{ik\_llama.cpp} reaches 2.16 and
2.48~tokens/s at the 24 and 32~GiB cache-budget points, respectively.
Under total-memory accounting, Dynamic Avg2.5 uses approximately
29.2~GiB and reaches 8.89~tokens/s, \(3.58\times\) the throughput of
\texttt{ik\_llama.cpp} under its 33.27~GiB limit. This is a
complete-runtime comparison rather than an isolation of any individual
mechanism.

Figure~\ref{fig:e2e-throughput}(b) shows the same progression on
GLM5.2. Dynamic Avg4 reaches 1.54~tokens/s, close to the
1.58~tokens/s of Uniform4, while Dynamic Avg2.5 reaches
2.64~tokens/s, \(1.67\times\) that of Uniform4.
\texttt{ik\_llama.cpp} reaches 0.72 and 0.77~tokens/s at the
24 and 32~GiB cache-budget points, respectively.

\subsection{Separating Fewer Requested Bits from Cross-Precision Reuse}
\label{sec:eval-cross-precision}

Dynamic Avg2.5 combines lower miss payloads and cache footprint with
cross-precision reuse. To isolate the latter, we replay identical
requests and bit-width assignments under otherwise identical runtimes.
The Independent configuration caches each \((e,b)\) pair separately and reloads its
complete prefix on a miss, whereas PagedMoE shares resident bitplanes
and fetches only the missing suffix.

\begin{figure}[t]
\centering
\begin{minipage}[t]{0.49\linewidth}
\centering
\includegraphics[width=\linewidth, trim=0 1mm 0 3mm, clip]
{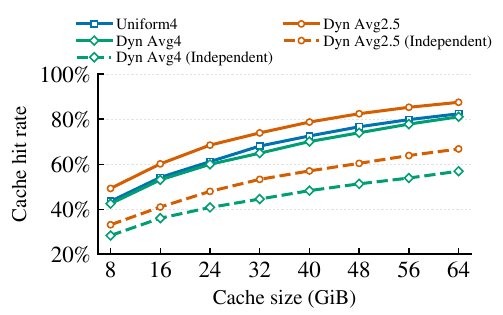}
{\small (a) Byte hit rate.}
\end{minipage}
\hfill
\begin{minipage}[t]{0.49\linewidth}
\centering
\includegraphics[width=\linewidth, trim=0 1mm 0 3mm, clip]
{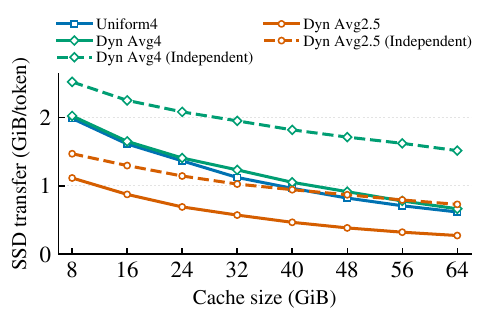}
{\small (b) SSD traffic.}
\end{minipage}
\caption{Effects of cross-precision reuse under identical expert requests
and bit-width assignments on Qwen3.5-397B-A17B.}
\Description{Two line charts comparing independent precision states and
cross-precision reuse in terms of byte hit rate and issued SSD read bytes
per generated token.}
\label{fig:reuse-locality-traffic}
\end{figure}

Figure~\ref{fig:reuse-locality-traffic} compares Independent Precision with
PagedMoE's shared state under identical requests and assignments. The
hit-rate values differ from Figure~\ref{fig:token-level-precision}(d),
which reports exact-object rather than byte hit rates.

At 24~GiB, cross-precision reuse increases the Dynamic Avg4 and
Dynamic Avg2.5 byte hit rates from 40.81\% to 59.98\% and from
47.94\% to 68.50\%, respectively. Their SSD read traffic correspondingly
decreases from 2.08 to 1.41~GiB/token and from 1.14 to
0.69~GiB/token. Because cache identity is the only difference, these
results isolate the I/O benefit of cross-precision reuse. 

\begin{figure}[t]
\centering
\begin{minipage}[t]{0.485\linewidth}
\centering
\includegraphics[width=\linewidth, trim=0 1mm 0 1.5mm, clip]
{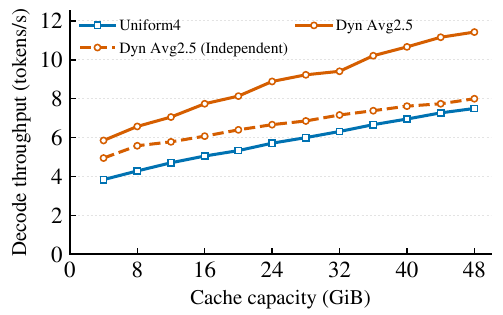}
{\small (a) Qwen3.5-397B-A17B.}
\end{minipage}%
\hfill
\begin{minipage}[t]{0.485\linewidth}
\centering
\includegraphics[width=\linewidth, trim=0 1mm 0 1.5mm, clip]
{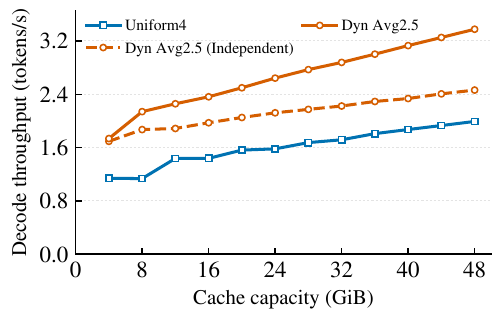}
{\small (b) GLM5.2.}
\end{minipage}
\caption{
End-to-end decomposition of fewer requested bits and
cross-precision reuse on (a) Qwen3.5-397B-A17B and (b) GLM5.2.
For each model, the independent and shared Dynamic Avg2.5
configurations replay identical expert requests and bit-width
assignments.
}
\Description{
Two line charts comparing Uniform4, Dynamic Avg2.5 with independent
precision states, and Dynamic Avg2.5 with PagedMoE's shared
progressive cache on Qwen3.5-397B-A17B and GLM5.2.
}
\label{fig:cross-precision-throughput}
\end{figure}

Figure~\ref{fig:cross-precision-throughput} shows how the reduction in
SSD traffic translates into end-to-end throughput. At 24~GiB,
Uniform4 and Dynamic Avg2.5 (Independent) reach 5.71 and
6.67~tokens/s, respectively, so fewer requested bits alone provide a
\(1.17\times\) improvement. With the same assignments, PagedMoE's shared state reaches
8.89~tokens/s, \(1.33\times\) the throughput of Independent and
\(1.56\times\) that of Uniform4. Across cache capacities on
Qwen3.5-397B-A17B, PagedMoE achieves
\(1.18\times\)--\(1.44\times\) the throughput of Independent. 

GLM5.2 shows the same decomposition: At 24~GiB, Uniform4, Dynamic Avg2.5 (Independent), and
PagedMoE reach 1.58, 2.12, and 2.64~tokens/s, respectively. These
correspond to a \(1.34\times\) improvement from fewer requested bits
and an additional \(1.25\times\) from cross-precision reuse, for a
combined \(1.67\times\) improvement. Together, these results show that
fewer requested bits reduce expert-data demand, while the shared
progressive state prevents cache fragmentation and rereading of
overlapping data across precision changes.

\subsection{Design Validation and Runtime Costs}
\label{sec:eval-conditions}

Having established the benefit of shared progressive state, we next
evaluate its representation cost, pre-I/O allocation, direct
execution, and explicit management relative to mmap.

\begin{figure}[t]
\centering

\begin{minipage}[t]{0.485\linewidth}
\centering
\includegraphics[width=\linewidth, trim=0 1mm 0 1.5mm, clip]
{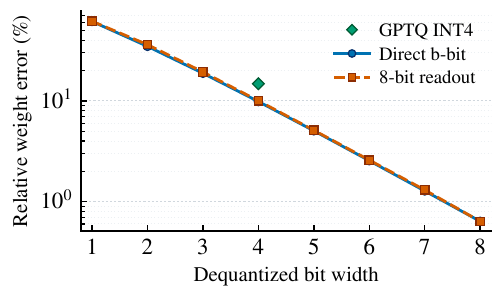}
\parbox[t]{\linewidth}{
    \centering\footnotesize
    (a) Prefix-compatible quality.
}
\end{minipage}%
\hfill
\begin{minipage}[t]{0.485\linewidth}
\centering
\includegraphics[width=\linewidth]
{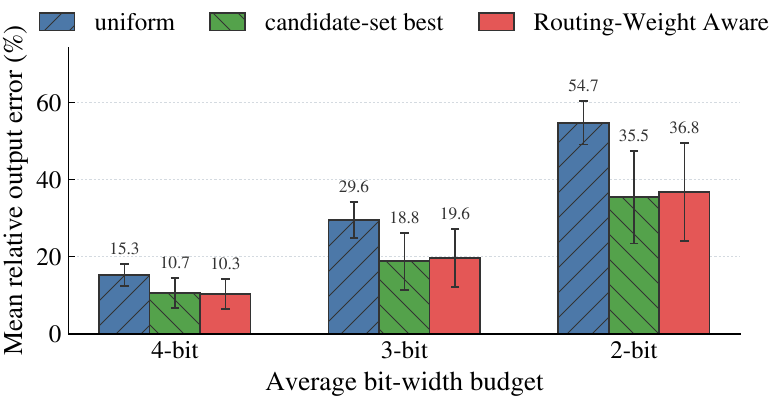}
\parbox[t]{\linewidth}{
    \centering\footnotesize
    (b) Pre-I/O allocation.
}
\end{minipage}

\vspace{3pt}

\begin{minipage}[t]{0.485\linewidth}
\centering
\includegraphics[width=\linewidth]
{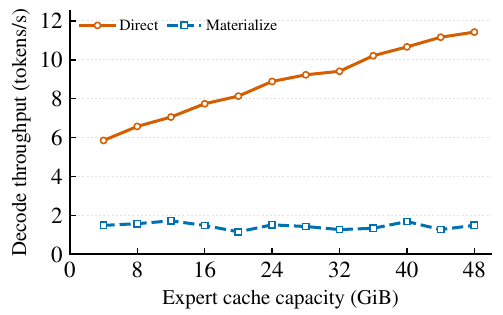}
\parbox[t]{\linewidth}{
    \centering\footnotesize
    (c) Direct execution.
}
\end{minipage}%
\hfill
\begin{minipage}[t]{0.485\linewidth}
\centering
\includegraphics[width=\linewidth]
{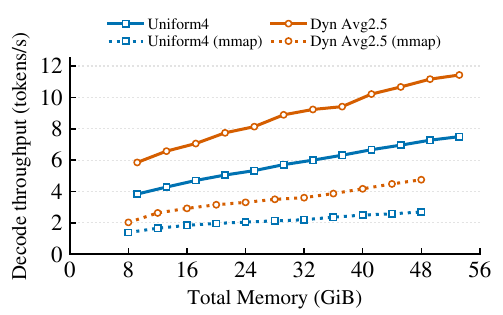}
\parbox[t]{\linewidth}{
    \centering\footnotesize
    (d) Explicit management vs.\ mmap.
}
\end{minipage}

\caption{
Necessary conditions and costs of PagedMoE's progressive-state design:
(a) prefix-compatible reconstruction quality;
(b) pre-I/O allocation under equal average-bit budgets;
(c) direct execution versus BF16 materialization; and
(d) explicitly managed versus mmap-backed runtimes under total-memory
accounting.
}
\Description{
Four panels evaluating the necessary conditions and costs of PagedMoE's
progressive-state design: prefix-compatible reconstruction quality,
pre-I/O precision allocation, direct compact execution, and explicitly
managed versus mmap-backed runtimes.
}
\label{fig:eval-conditions}
\end{figure}

\subsubsection{Cost of Prefix Compatibility}
\label{sec:eval-quantizer-quality}

As quantified in Section~\ref{sec:progressive-representation},
PagedMoE's 8-bit index payload is twice that of a fixed-4-bit
representation, but one copy exposes all 1--8-bit levels instead of the
36 bits per weight required by independent variants; tree-codebook
metadata adds less than 0.8\%.

Figure~\ref{fig:eval-conditions}(a) shows that independently trained
and progressive-readout quantizers nearly overlap across bit-widths
for the routed-expert weights in layer~30 of Qwen3.5-397B-A17B,
indicating little additional reconstruction error for prefix
compatibility. At four bits, PagedMoE yields 10.02\% relative weight
error, compared with approximately 14.8\% for GPTQ-INT4. Because the
formats differ, GPTQ-INT4 serves only as a practical reference rather
than evidence of general superiority.

\subsubsection{Effectiveness of Pre-I/O Precision Allocation}
\label{sec:eval-allocation}

Figure~\ref{fig:eval-conditions}(b) compares three policies under the
same prefix-compatible representation and average-bit budget:
Uniform assigns the same precision to all selected experts; the
restricted candidate-set best selects among predefined schedules using
BF16 outputs unavailable at runtime; and PagedMoE uses routing weights
and offline expert-specific error profiles.

At average budgets of four, three, and two bits, Uniform produces
relative routed-output errors of 15.3\%, 29.6\%, and 54.7\%, whereas
PagedMoE reduces them to 10.3\%, 19.6\%, and 36.8\%, close to the
restricted candidate-set best at 10.7\%, 18.8\%, and 35.5\%.
PagedMoE's slightly lower error at four bits does not imply outperforming
an oracle, because the reference searches only a restricted set of
predefined schedules.

\begin{table}[t]
\centering
\footnotesize
\setlength{\tabcolsep}{3.5pt}
\renewcommand{\arraystretch}{0.95}
\caption{WikiText-2 perplexity under different precision-allocation
policies on Qwen3.5-397B-A17B. Lower is better.}
\label{tab:qwen-allocation-ppl}
\begin{tabular}{@{}lccc@{}}
\toprule
\textbf{Policy} & \textbf{Rank-to-bit rule} &
\textbf{Avg. bits} & \textbf{PPL} \\
\midrule
Dynamic Avg2.5 & PagedMoE allocator  & 2.5 & \textbf{3.240} \\
Rank-Based-A & top-5: 4, others: 1 & 2.5 & 3.304 \\
Rank-Based-B & top-5: 3, others: 2 & 2.5 & 3.334 \\
Rank-Based-C & top-3: 4, others: 2 & 2.6 & 3.282 \\
\bottomrule
\end{tabular}
\end{table}

Table~\ref{tab:qwen-allocation-ppl} compares PagedMoE with two-level
routing-rank policies end to end. At the same 2.5-bit budget, Dynamic
Avg2.5 achieves lower perplexity than both Rank-Based Avg2.5 policies and 
outperforms Rank-Based Avg2.6 despite using fewer average bits. Together,
the routed-output and perplexity results show that PagedMoE's
deployable pre-I/O allocator improves the quality--bit tradeoff over
uniform precision and simple rank-based allocation.

\subsubsection{Preserving Compactness through Direct Execution}
\label{sec:eval-compact-execution}

After loading a requested prefix, the runtime must consume it directly;
materializing a full-precision expert would reintroduce DRAM traffic.
Figure~\ref{fig:eval-conditions}(c) compares \emph{Direct}, which
reconstructs small weight tiles from cached bitplanes and immediately
consumes them in fused GEMV, with \emph{Materialize}, which constructs
a complete BF16 expert matrix before BF16 GEMV. Both paths replay
identical expert requests, precision assignments, and cache and I/O
decisions.

Direct achieves \(3.94\times\)--\(8.79\times\) the decode throughput
of Materialize by avoiding the materialization, writeback, and
rereading of complete BF16 expert matrices. This result confirms that
the progressive state must remain compact through CPU execution.

\subsubsection{Complete-Runtime Comparison with mmap}
\label{sec:eval-explicit-runtime}

Figure~\ref{fig:eval-conditions}(d) compares PagedMoE with an
mmap-backed runtime under total-memory accounting. Both use the same
progressive checkpoint, bitplane layout, and precision assignments, so
mmap preserves cross-precision reuse through the operating-system page
cache. The mmap-backed runtime delegates residency, replacement, and SSD fetching
to demand paging, whereas PagedMoE manages prefix residency and
missing-suffix reads explicitly.

At a 32~GiB limit, mmap-backed Uniform4 and Dynamic Avg2.5 reach
2.20 and 3.61~tokens/s, respectively. The \(1.64\times\) improvement
shows that progressive sharing and lower-bit allocation remain
effective even under mmap. With a 24~GiB bitplane cache and approximately 29.2~GiB total
memory, the corresponding PagedMoE configurations reach 5.71 and
8.89~tokens/s, demonstrating the benefit of explicit management. At comparable total memory, mmap-backed Dynamic Avg2.5 reaches
3.61~tokens/s at 32~GiB, compared with 2.48~tokens/s for
\texttt{ik\_llama.cpp} under its 33.27~GiB limit. This shows that
PagedMoE's advantage over \texttt{ik\_llama.cpp} does not arise from
explicit management alone. The PagedMoE--mmap gap remains a complete-runtime comparison
because residency, replacement, I/O, and scheduling change jointly.

\section{Related Work}

\paragraph{MoE offloading and caching.}
Existing systems optimize expert placement, caching, prefetching, and heterogeneous CPU--GPU execution~\cite{moeinfinity,fiddler,ktransformers}. They typically manage each expert as a fixed-format object. PagedMoE instead manages its progressively refinable precision state, enabling cross-token reuse across precision changes.

\paragraph{Progressive multi-precision quantization.}
Any-Precision LLM and MatQuant show that multiple bit widths can share
a nested representation~\cite{park2024anyprecision,nair2025matquant}.
PagedMoE applies this principle to SSD-resident MoE inference, using
the shared representation as persistent expert state across SSD
storage, DRAM caching, and execution.

\paragraph{Dynamic-precision MoE inference.}
Offline methods such as MxMoE and GEMQ assign expert or block precisions using profiled sensitivity and activation statistics~\cite{mxmoe,gemq}. HOBBIT serves less-critical cache misses at low precision, while D$^{2}$MoE combines token-adaptive bit-width selection with nested weights and bit-width-aware I/O scheduling~\cite{hobbit,d2moe}. SliceMoE and DyMoE use gating scores or runtime importance for mixed-precision caching and scheduling~\cite{slicemoe,dymoe}. PagedMoE extends this direction to fine-grained 1--8-bit expert precision for SSD-resident inference. Its pre-I/O allocator combines current routing weights with offline expert- and precision-specific approximation errors, providing an explicit quality safeguard under a fixed average-bit budget. PagedMoE further integrates a DRAM bitplane cache with direct CPU execution, allowing cached quantized states to be reused across tokens and consumed without materializing BF16 experts.

\paragraph{Compact quantized execution.}
Prior fused and lookup-based kernels execute low-bit weights without fully materializing them~\cite{lutgemm,flute,tmac}. PagedMoE uses a fused kernel to consume the selected bitplane prefix directly; this is an execution enabler rather than its primary novelty.

\section{Conclusion}

PagedMoE shows that token-level dynamic precision benefits
SSD-resident MoE inference only when precision changes preserve cached
expert state. It therefore represents each expert as one progressively
refinable state: after routing, a pre-I/O allocator selects the required
prefix, the runtime reuses or fetches only its missing suffix, and a
fused CPU kernel executes the compact prefix directly.

At an average routed-expert precision of 2.5 bits, PagedMoE achieves
quality close to GPTQ-INT4 on Qwen3.5-397B-A17B and to MXFP4 on most evaluated GLM5.2 tasks. With a 24~GiB expert
cache, it reaches 8.89~tokens/s on Qwen3.5, \(1.33\times\) the throughput of the
same requests with independent precision states and \(1.56\times\)
that of uniform 4-bit execution. These results show that fewer
requested bits and cross-precision reuse are complementary: one
reduces expert-data demand, while the other prevents precision changes
from breaking reuse.

\bibliographystyle{ACM-Reference-Format}
\bibliography{refs}

\end{document}